\documentstyle[12pt,aasms4]{article}
\received{}
\revised{}
\accepted{}

\cpright{}{}
\journalid{}{}
\articleid{}{}
\paperid{}

\cpright{}{}
\ccc{}

\lefthead{Grandpierre}
\righthead{ }

\begin{document}

\title{ A Dynamic Solar Core Model: the SSM-like solution}

\author{Attila Grandpierre}

\affil{Konkoly Observatory, P.O. ~Box ~67, H--1525,
Budapest, Hungary \\
electronic mail: grandp@ogyalla.konkoly.hu}

\authoraddr{Konkoly Observatory, P.O. ~Box ~67, H--1525,
Budapest, Hungary}

\vskip 10cm

\begin{abstract}
I point out that all the arguments available in the literature against an
astrophysical solution of the solar neutrino problem(s) do not exclude a yet not recognised class of solar models, in which an explosive energy source is present in the solar core besides the standard proton-proton and CNO cycle. It is shown from first principle physics that stars have a non-equilibrium energy source as well, arising from transient local thermonuclear runaways. I derive a model independent inequality, which shows that the problem of the missing beryllium neutrinos lies in that the SuperKamiokande contains a term arising from neutrinos of a non-pp,CNO source. Recent indications suggest strongly that this term plays a more significant role in the solar neutrino problem than neutrino oscillations. The runaways can produce high-energy electrons and high-energy axions, and may produce muon and tau neutrinos above $10^{11}$ K, and electron antineutrinos. I show that subtracting the runaway term from the SuperKamiokande results the remaining data are compatible with each other, within the frame of a generalised luminosity constraint and standard neutrinos. Allowing non-pp,CNO reaction chains a new approach arises to interpret the neutrino detector data, opening new horizons to solar neutrino astrophysics. I point out, that the temperature dependence of the individual neutrino fluxes is related to pure nuclear physics, but the usual luminosity constraint is model dependent and actually is a questionable assumption. The explicit temperature dependence leads to $\Phi_{pp} \propto T^4$ for the more general case instead of the usual $\Phi_{pp} \propto T^{-1/2}$ for the SSM luminosity constraint.
 
I assume a Sun analogue to the SSM, in the sense that the actual, average temperature can be described by a single parameter, in terms of the SSM temperature. This is the "SSM-like solution" of the neutrino-flux equations.  The separate neutrino detector equations lead to separate detector-related temperatures with the neutrino detector rates. The results show a slightly lower than standard central temperature. I attempt to show that helioseismology is not in a necessary conflict with the dynamic solar model presented here. The results of the presented calculations may propose solutions to the problems of solar and atmospheric neutrino oscillations without an ad hoc introduction of sterile neutrinos and present predictions to Borexino and SNO measurements.


\end{abstract}
%
%
\section{Introduction}
In the last thirty years the solar neutrino problem(s) proved to be one of the most long-standing puzzles of present-day physics and astrophysics (see the reviews [1] - [4]. There are not only one sources of neutrino problems, since they involve the atmospheric neutrino anomaly [5], the three solar neutrino problems of the Standard Solar Model (SSM, see e.g. [6], [2]), the anomalous results of the Liquid Scintillator Neutrino Detector (LSND) (see e.g. [7]), but the supernova neutrinos [8] and the cosmological context [9] also contributes to the evaluation of the neutrino problems. I find it more reasonable to pay attention to solar astrophysics when attempting to understand the solar neutrino problems, than to look for the answer exclusively in the domain of particle physics. I suggest that we can subtract useful input physics from astrophysics as well.

The seemingly contradicting results reached in the above fields have been generated numerous papers and approaches. Different attempts were invented to resolve the apparent paradox. Now it seems that the paradox may be resolved only if one or more type of sterile neutrinos are introduced [7], [10]-[12], which would originate only in neutrino oscillations, and would not interact even weakly with any other kind of matter. Even in the hypothetical case, if the future neutrino experiments will detect just the values expected for the MSW effect, it would not solve the physical neutrino problems (of atmospheric and LSND anomalies) without sterile neutrinos. In that case, the theoretical and experimental determination of the characteristic properties of the sterile neutrinos would be the only way to solve the neutrino problems - but this approach needs a long time to test. In contrast to these attempts, I introduce here an approach based on first principles and physically more robust facts. This new approach - the dynamical solar core approach - is based on the recognition that a fundamental factor, an explosive energy source of the Sun is not taken into account in our standard picture of the Sun. Also, I suggest paying more attention to the largely ignored factors, namely, the solar core-related astrophysical problems. I attempt to show here, that these ignored factors could present useful and non-negligible inputs to neutrino theories. A consequence of the non-pp,CNO nuclear reactions is that the different type of neutrino detectors show different sensitivities to these new ingredients of the solar energy production, and therefore a careful analysis is necessary to subtract their contribution from the overall values. 

The problem is that the neutrino detectors measure certain events, and we need a relation between the events and the individual neutrino fluxes. But to determine which relation is at work, we would need the energy distribution of the neutrino flux "a priori", because this energy distribution is the one which determines the relation between the incoming neutrino fluxes and the measured events. If non-standard nuclear reactions produce also neutrinos, it is not evident how to find the necessary relation between the neutrino fluxes and the observed counts. In this paper I attempt to determine the contributions of the non-pp,CNO neutrino fluxes to the individual neutrino detectors. The solution of the standard pp,CNO neutrino flux equations show clearly that they require lower than SSM neutrino-related temperatures. A conjecture is that in this case an additional energy source has to be at work in the solar core. This conjecture is already indicated by first principle physics and by a significant list of the astrophysical problems concerning the solar core [13]. Taking into account a non-pp,CNO energy source, a part of the fluxes measured by Homestake, Gallex and SuperKamiokande is produced by the high-energy electron neutrinos and axions, and, possibly, by the muon and tau neutrinos produced above $10^{11} K$ by the explosive reactions, so one has to diminish the observed values with the proper amounts to be calculated. At the same time, the presence of a new energy source allows to lower the central temperature of the Sun without violating the stationary Sun assumption. $\Delta T \simeq 6 \%$ would lead to the measured $pp$, $Be$ and $B$ neutrino fluxes, while - as I attempt to show in the followings - the dynamical solar core model seem to be in principle consistent with the helioseismic results. In that way, a "distorted" energy spectrum arises for the solar neutrinos, as compared to the standard solar model of [14], not because of neutrino oscillations but by explosive processes occurring in the solar core. The presence of non-pp,CNO energy source allows us to escape from the "strongest argument for a particle physics solution to the solar neutrino problem" which "arises from energy conservation (the luminosity constraint)" [15], and suggest that a more general luminosity constraint characterises the real Sun.

From the works of [6], [16]-[23], it is clear that the neutrino flux equations of the different type of neutrino detectors contradict to each other, even when taken any pair of them together, within the frame of standard neutrinos and the SSM luminosity constraint. The arguments against an astrophysical solution are based on the rough values of the neutrino detectors that are contradicting to each other assuming standard neutrinos. As Heeger and Robertson [20] wrote, "the least model-dependent questions that can be asked are, is it possible to describe the neutrino spoectrum with any combination of the known sources in hydrogen burning? Is the total neutrino flux consistent with the solar luminosity?" My point is to show that these two questions do not necessarily refer to the same content, when a non-pp,CNO energy source is present. In that case, the use of the SSM luminosity constraint is not allowed. Similarly. the "Last Hope" of the astrophysical solutions [21] allows arbitrary parameterisation of the neutrino fluxes and still leads to the exclusion of the astrophysical solutions. This is due to the fact that they also did not allow non-standard energy sources and related non-standard neutrino fluxes in their equations, and that they used the rough neutrino detector data, which are contradicting with standard neutrinos. In contrast, the model independent inequality which I derive here clearly shows that the root of the problem of the missing beryllium neutrinos goes back to the fact that the SuperKamiokande has to detect non-pp,CNO neutrinos as well. Other facts - the sensitivity of SuperKamiokande to neutral currents, axions, anti-neutrinos etc. - suggest strongly the presence of non-pp,CNO electron neutrinos even without neutrino oscillations. The Cl-Ga comparison also favours more strength at higher energies [20]. 

Some indications already were recognised in the published literature which points toward the approach developed here. Calabresu et al. [24] noted that as the observed Kamiokande neutrino flux $\Phi_B$ being just a factor two below the SSM (while the other detectors show a higher neutrino deficiency), it looks that experiments are observing too high $\Phi_B$. On the other hand, Baltz, Goldhaber and Goldhaber [25] already assumed that the neutrino deficiencies found are partially due to an overestimate of the emission intensity of the $^8B$ neutrinos by an amount of $36 \%$. Cleveland et al. [26] recently remarked that the difference between the data of the Homestake and Kamiokande detectors has to arise largely by the non-electron neutrino flux, which are not detected by the Homestake but are measured by the SuperKamiokande detector. 

%
%
\section {Astrophysical problems of the solar core }
Such tantalising astrophysical problems became known like the anomalous heavy element enhancements of flares (involving e.g. Fe,  which can be produced only above $10^9$ K, the absolute enhancement factor extends from 4 to 20 and larger [27]-[28]), the nitrogen-enigma (indicating the production of $^{15}N$, which can be produced only above $10^8 K$, [29]), the anomalously slow solar core rotation (the best solar model of Pinsonneault et al. [30]) suggests 4 to 15 times the observed core rotation rate), the anomalous heavy element enhancement of the solar corona ([13], and the references therein), and - without the aim of completion - the rigid rotation of the activity generating centres in the differentially rotating photosphere, indicating a deep origin from a rigid rotating core [31].

One has to take care with the statement published frequently in the literature that astrophysical solutions cannot work solving the solar neutrino problems [17]-[24], [32]. When these authors abandoned the SSM luminosity constraint, they did so while keeping the neutrino data at their face values. And since the rough neutrino data are contradicting to each other, therefore the result they derived was not compatible with a standard solution to the neutrino problems. I have to note here, that all these results refer only to hydrogen-based $ \it slow$ standard fusion reactions, therefore their consequences would not apply to the case considered here in which a $ \it rapid $ energy source also contributes to the energy production and supports a part of the solar luminosity. Heeger and Robertson [20] stated that even exotic energy sources, which produce energy but no neutrinos, are excluded for solving the solar neutrino problems. But their argument is based on the fact that the neutrino detector data contradict to the SSM solution of the solar neutrino problem even when abandoning the luminosity constraint. The point suggested here is a step ahead towards model independence: to abandon the SSM luminosity constraint and replace it with a more general one, AND ALSO to abandon a fundamental assumption behind the SSM. This assumption to abandon is that the energy produced in the proton-proton cycle is $ \simeq 98 \%$ and in the CNO cycle it is $ \simeq 2 \%$ of the solar energy production. I suggest that a new, non-pp,CNO ingredient also participates in the energy production. The contributions of the new ingredient to the detectors have to handle separately. In this way, one has to find the standard part of the neutrino detector data and use these values in the standard neutrino flux equations. 

There are presented only some estimation regarding the formation, development, energetics, mass, life-time, and surface connection of the hot bubbles [13]. Nevertheless, I have to point out here that the travel of the "hot spots" from the Earth's core through the mantle to the volcanic sites at the Earth's surface seems to be a much more significant achievement than the travel of hot bubbles from the solar core towards the surface - yet the basic physics may show significant similarities. 

Cumming and Haxton [33] and Haxton [34] present another type of a yet un-eliminated astrophysical solution. Most recently, Du Jiulin [35] has calculated that if a flow velocity field is present in the solar core, the proton-proton reaction system becomes unstable and switches into an other energy producing channel, which leads to a local $^3He$ enhancement. The result of Du Jiulin [35] (1998) offers a connection between the macroscopic motions as described by the present author [13], and the $^3He$ enhancement generating preferential $^7Be$ flux depression over the $^8B$ one as suggested by [33] for the astrophysical solution of the beryllium-neutrino problem.    

%
%
\section { The present allowed parameter ranges of the MSW mechanism }

The recent measurements of SuperKamiokande present the yet strongest indication that at least one kind of neutrinos (the muon-neutrino) has a finite mass. The experimental results lead to incompatible ranges of assumed neutrino masses and mixing angles within the frames of the MSW effect when allowing three flavours. Now it is clear that there is no such a purely physical three-flavour neutrino theory that satisfies the atmospheric, the LSND and the solar neutrino problems.

Ten years ago Bahcall wrote that the MSW mechanism is attractive since it does not need fine tuning, as it works well for variations by orders of magnitude in the $(sin^2 2 \theta, \delta m^2)$ parameter space [36]. In contrast to this, at present the allowed region decreased to a point-like area. Paterno and Scalia [37] refer to this fact as "hypotheses on non-conventional neutrino properties are strongly disfavoured, except for matter neutrino oscillations, the latter surviving within very narrow limits". Similarly, Narayan et al. [38] remarked that "there is very little leeway in the allowed values of $\delta m^2$ and $sin^2 2\theta$".

Bahcall, Krastev and Smirnov [39] found that the best fitting values of the solar neutrino oscillations have only a low confidence, with a level around $ 7 \%$. This fit is obtained when not only the total rates of the neutrino detectors, but the SuperKamiokande electron recoil energy spectrum are also taken into account. When the constraints from the SuperKamiokande zenith-angle distribution is included together with the constraints from the four measured total rates and the SuperKamiokande electron recoil energy spectrum, the best-fit MSW solution is acceptable with a confidence level of $7 \%$. Only the small mixing angle solution survives at the $99 \% C.L.$. The large mixing angle and the LOW solution are marginally ruled out. By my judgement, this fit cannot be regarded anything else but a bad one. Moreover, the cosmological context seems to rule out just the small angle MSW explanation. Regarding the cosmological neutrino problem, Georgi and Glashow [40] pointed out that recent data, leading to nearly maximal mixing, rule out the small-angle MSW explanation of solar neutrino observations, if relic neutrinos comprise at least one percent of the critical mass density of the universe. 

I suggest that the reason is that the solar electron-neutrino oscillations do not offer a complete solution to the solar neutrino problems, since other factors (also) may play significant roles. Taking into account the non-pp,CNO chains it would be possible to reach a solution to all of the neutrino problems with three flavours only, and, possibly, improving the MSW fit. The approach I will outline here consists in evaluating the newly arisen possibilities, to explore some conjectures of the presence of explosive energy sources in stellar cores. 

%
%
\section { Fundamental thermonuclear inequilibrium of stellar cores }

The Sun is believed to be in thermodynamical equilibrium and to generate its energy by reactions building up helium from hydrogen exclusively by the ordinary pp,CNO reactions.  I emphasise that, in contrary, stellar energy producing cores are necessarily in an unstable state. This instability arises in the form of thermonuclear runaways (see [41]-[42], [13]). These runaways are generated by the ultrasensitive character of the thermonuclear energy production. Let us see how it develops.

The ultrasensitive character of the thermonuclear energy production to variations in the temperature is known since a long time. This ultrasensitivity explains the mass-luminosity (M-L) relation $M \propto L^{3.5}$ which formula shows that the larger the mass of a star, the much higher is its luminosity and therefore its energy production per gram per unit of time. This means that the energy producing cores of the stars are exponentially sensitive to the temperature. Actually, the energy production in stellar cores goes with a high power of temperature, from around $4$ for hydrogen-burning stars to a power of $ n > 20$ for helium and heavy element burning stars. This extreme temperature dependency do not lead to a global eruptive instability of the star since the self-gravity of the star leads to a negative heat capacity for a global expansion of the star. 
Nevertheless, this argument can not apply to $\it local $ temperature perturbations, since any small $\it local $ volume of the star has a positive heat capacity, therefore all the local regions in which a temperature perturbation arise by some circumstance, produce a kind of decaying (or not decaying) thermonuclear runaway. 
Because of the high temperature dependence, even microscopic processes, like stochastic heat fluctuations will enhance the rate of the nuclear reactions in their environment. The enhancement will lead to a further increase of enhancement and a simultaneous enhancement of dissipation. When the temperature enhancement do not extend to a larger-than-critical size, the perturbations will not result to a kind of convective instability, and to the large-scale transport of the hot bubble outwards. Nevertheless, the perturbations do not necessarily decay to zero, since their existence leads to a self-regenerative sustainment of the thermal inequilibrium, which could be characterised with a double temperature distribution both for electrons and for protons, similar to the surface of a temperature-sea under a permanent storm. One curve can characterise the covering envelope of the tops of the temperature waves, or the average height of the waves, and another one could describe the standard "sea level".

Another effect arises from the fundamental stellar inequilibrium. 
I proposed that the interaction of tidal waves of the planets with a velocity ${\bf v}$ and the global magnetic field of the Sun ${\bf B}$ induces an electric heating ${\bf E}$ in the solar core ${\bf E} = 1/c \ ({\bf v} \times {\bf B})$ in a form of a current filament [42]. 
For a typical tidal flow velocity at the solar core $100 cm/s$ [43], and with a magnetic field strength around $10^6 G$ (much lower ${\bf B}$ also works to initiate the runaway), the generated electric field will be 1 V/cm. 
Electric fields of a similar strength are present in the electric discharges observed in the flare phenomena of the outer solar atmosphere. The generated electric field in the solar core accelerates the charged particles to a high speed. The acceleration length $l(a)$ is $U/eE$, where $U$ is the energy of the particle after the acceleration. Since the mean free path of the protons at the solar core is around $5 \times 10^{-4} cm$, the electric field will heat the solar plasma locally to very high temperatures. For $U=200 keV$ the related kinetic temperature is $3 \times 10^9$ K. At the peak of the He-flash the temperature is only $3 \times 10^8 K$ [44]. The most important factor determining the development of the local temperature enhancement will be the time-scale of the thermonuclear runaway. This can be derived from the energy equation
\begin{eqnarray}
cdT/dt = \epsilon_{pp} = \rho x_1^2 \epsilon_0 (T/T_0)^{\nu}
\end{eqnarray}
This equation is solved with the substitution of a new variable $T^{1-\nu}$. 
The solution is
\begin{eqnarray}
T_2 = T_1/({1-( \rho x_1^2 (\epsilon_0 (\nu - 1)/C_vT_1))t})^{(1/(\nu -1))}
\end{eqnarray}
and the time-scale of the thermonuclear runaway is
\begin{eqnarray}
\tau_r = 3/2(RT)( \rho x_1^2 \epsilon_0 (\nu - 1)).
\end{eqnarray}
Formula (3) gives this scale for the slowly changing, standard solar core model a value around $3 \times 10^7 years$. For the explosive nuclear reactions, with $\epsilon_{pp} = 5 \times 10^{17} ergs \ g^{-1} s^{-1}$ [45], $\tau_r$ will be around $6 \times 10^{-5} s$. At the same time, the time scale for the expansion, estimating as being around the hydrodynamic time scale for free expansion, is 
\begin{eqnarray}
\tau_{exp} = {1/ \rho (d \rho / dt)},
\end{eqnarray}
which formula tells us that $\tau_{exp} \simeq (24 \pi G \rho)^{1/2} = 1/446\rho^{-1/2} s^{-1} $
[46]. Therefore, at the solar centre, the time scale for expansion is around $1/40 s$, i.e. much larger than that of the thermonuclear runaway. This means that the local thermal expansion cannot compensate the heating arising by the thermonuclear runaway, i.e. a thermonuclear runaway may set up.  
The growth of the region heated by the electric current filament may or may not reach the critical size determined by the relation of the actual to the critical Rayleigh number, depending on ${\bf v}, {\bf B}$ and the coefficients parameterising the environment. When the boistering volume does not reach the critical size [47], the instability do not lead to macroscopic motions but to thermal inequilibrium. Above the threshold the local hot regions may rise so that the inner energy production surplus surmounts the dissipative effects. 
In this case, the microinstabilities evolve to macroinstabilities, to the rise 
of 'hot bubbles' towards the surface of the star. I mention, that Zel'dovich, 
Blinnikov and Sakura [48], without drawing any astrophysical consequence, independently derived a calculation showing that every star has to be in a permanent local thermodynamic equilibrium. Since their argument is not available in English, therefore I present it in the Appendix for the interested readers.

%
%
\section{ Helioseismology }

Helioseismic data also indicate the presence of macroscopic motions in the solar interior [49]. I suggest that the deviancies of the SSM to the helioseismically derived sound-speed is connected to a non-standard energy-source of the Sun. Although the different groups GONG, BISON, GOLF, IRIS reached results somewhat contradicting to each other, they agree to find that the rotation velocity of the inner solar core is between 300-600 nHz [50], while the surface rotation rate is around 420 to 450 nHz. This means that the solar core rotates slightly slower or faster than the surface. This is in strict contradiction with the 'best solar model' [30] which shows that the rotation rate of the inner solar core at $ r < 0.2 R_{Sun}$ should be from $4$ to $15$ times faster than the surface rate. The slow rotation rate of the solar core indicates that a dynamic process coupling the solar core to the surface was not taken into account. 

It is questioned ([13], [34], [51]) the relevancy of the helioseismic inversions to deduce the central solar temperatures. The problem is similar to the interpretation problem of the counts of the neutrino detectors: to interpret the observed oscillation frequencies, we need an "a priori" solar model that supplies the numerical background to the inversion that it can become possible. If this "background solar model" is not the one realised by the Sun, it is questionable the reality of any deduction based on the SSM. Haxton regards the conclusion about the exclusion of mixed models to be premature [34]. As Primack [52] cites Haxton's personal communication, "claims that helioseismology data ruled out mixed models are now being reexamined, given that the original investigations did not use dynamically consistent solar models, but rather ad hoc perturbations of the SSM".

I have to add that magnetic fields, suggested to be produced (and expelled out quickly) in the solar core continuously, influence the solar model calculations in a way to increase the sound speed. It is not known, how such 3-dimensional effects, as the jet-like development of hot channels in which the hot bubbles are shot outward, modifies the helioseismic inversion process. In short, helioseismology is not able in its present form (see also in the Discussion section) to inform us unambiguously about the processes occurring in the innermost solar core: about its rotational rate, its magnetic field, its fundamental thermodynamic equilibrium.  

%
%
\section{ Basic equations }

\begin{eqnarray}
S_K = a_{K8} \Phi_8
\end{eqnarray}
\begin{eqnarray}
S_C = a_{C1} \Phi_1 + a_{C7} \Phi_7 + a_{C8} \Phi_8
\end{eqnarray}
\begin{eqnarray}
S_G = a_{G1} \Phi_1 + a_{G7} \Phi_7 + a_{G9} \Phi_8 ,
\end{eqnarray}
with a notation similar to that of Heeger and Robertson [20]: the subscripts i = 1, 7 and 8 refer to $pp + pep$, $Be + CNO$ and $B$ reactions. The $S_j$-s are the observed neutrino fluxes at the different neutrino detectors, in dimensionless units, j = K, C, G to the SuperKamiokande, chlorine, and gallium detectors. $\Phi_i$ are measured in $10^{10} \nu cm^{-2}s^{-1}$. Similar equations are presented by Castellani et al. [53], Calabresu et al. [24], and Dar and Shaviv [4] with slightly different parameter values.
Using these three detector-equations to determine the individual neutrino fluxes  $\Phi_i$, I derived that
\begin{eqnarray}
\Phi_8 = S_K/a_{K8}
\end{eqnarray}
\begin{eqnarray}
\Phi_1 = (a_{G7}S_C -a_{C7}S_G + S_K/a_{K8}(a_{C7}a_{G8} - a_{G7}a_{C8}))/D
\end{eqnarray}
\begin{eqnarray}
D = a_{G7}a_{C1} - a_{C7}a_{G1}
\end{eqnarray}
and
\begin{eqnarray}
\Phi_7 = (a_{G1}S_C - a_{C1}S_G + S_K/a_{K8}(a_{C1}a_{G8}-a_{G1}a_{C8}))/D'
\end{eqnarray}
\begin{eqnarray}
D'= a_{G1}a_{C7} - a_{C1}a_{G7}.
\end{eqnarray}\begin{eqnarray}\Phi_7 \times D' = a_{G1}S_C - a_{C1}S_G + (a_{C1} a_{G8}/a_{K8} - a_{G1} a_{C8}/a_{K8})S_K.
\end{eqnarray}
Obtaining these solutions, the root of the beryllium-problem goes to the circumstance that $D'> 0$, therefore the numerator has to be also positive. We know that the $\Phi_7$ can have only a physical, positive value. This fact requires that the following formulae has to be valid: 
\begin{eqnarray}
S_K < (a_{G1}S_C-a_{C1}S_G)/(a_{C1}a_{G8}/a_{K8}-a_{G1}a_{C8}/a_{K8})
\end{eqnarray}
Numerically,
\begin{eqnarray}
\Phi_7 = 0.4647 S_C -0.0014S_G - 0.5125S_K.
\end{eqnarray}
Now we see that the problem of the suppression of solar beryllium neutrinos is related to the circumstance  
that the coefficient of $S_C$ is smaller than  $S_K$, therefore, since all the $S_i$ values are positive, $\Phi_7$ cannot be physical. If we require a physical $\Phi_7$, with the numerical values of the detector sensitivity coefficients, 
this constraint will take the following form:
\begin{eqnarray}
S_K < 0.9024S_C - 0.0027S_G \simeq 2.115
\end{eqnarray}
Another general equation may be derived between the detector measurement parameters $S_i$ from the constraint that the $pp+pep$ neutrino flux has to be positive:
\begin{eqnarray}
S_K > 0.968S_C -0.030S_G \simeq 0.26,
\end{eqnarray}
with the observed values $S_K = 2.44$, $S_C = 2.56$, and $S_G = 72.2$ [14]. From these two inequalities, the inequality for the beryllium neutrino flux is
the one that is more constraining. This inequality of the neutrino detector rates was not derived previously in the published literature by my knowledge. This is a completely model independent inequality which shows that the SuperKamiokande (and possibly the other detectors as well) contains a term arising from neutrinos from a non-pp,CNO source. That is, with standard neutrinos, it is not physical to use the SuperKamiokande result $S_K$ at its face value $2.44$ in the standard neutrino equations. I have shown here that the root of the problem of the  missing beryllium neutrinos goes back to (14), (17). The detector rate inequality can be fulfilled only if we introduce an additional term $S_K(x)$ to represent the contribution of non-pp,CNO neutrinos to the SuperKamiokande measurements (and, possibly, if we modify the $S_C$, $S_K$ values properly). The presence of a non-electron neutrino term in the SuperKamiokande is interpreted until know as indication to neutrino oscillations. Nevertheless, thermal runaways are indicated to be present in the solar core that may produce high-energy electron neutrinos, as well as muon and tau neutrinos. Moreover, the explosive reactions have to produce high-energy axions to which also only the SuperKamiokande is sensitive. This indication suggests a possibility to interpret the neutrino data with standard neutrinos as well. 

What information can be subtracted from the neutrino flux equations about this extra term $S_K(x)$? To see this, I introduced our "a priori" knowledge on the pp,CNO chains, i.e. their temperature dependence. In this way one can derive the temperature in the solar core as seen by the different type of neutrino detectors. I note that the introduction of temperature dependence does not lead to solar model dependency. Instead, it points out the still remaining solar model dependencies of the previous SSM calculations and allowing other types of  chains, it removes a hypothetical limitation, and accepting the presence of explosive chains as well, it probably presents a better approach to the actual Sun.

%
%
\section{ Basic equations with temperature dependency: the SSM-like approach}

The dynamic model equation of the SuperKamiokande detector would be the following:
\begin{eqnarray}
S_K(T) = T^{24.5} \Phi_8(SSM) a_{K8},
\end{eqnarray}
But the calculations of the previous section has shown, that this SuperKamiokande-equation has to be completed with a new term
\begin{eqnarray}
S_K = T^{24.5} \Phi_8(SSM) a_{K8} + S_K(x),
\end{eqnarray}
where $T$ is the dimensionless temperature $T = T(actual)/T(SSM)$ and 
$S_K(x)$ is the contribution of additional, non-pp,CNO terms (like electron, moun and  tau neutrinos from the hot bubbles, axions, anti-neutrinos etc.) to the standard Kamiokande rate $S_K(0)$. 
\begin{eqnarray}
S_K(x) = S_K(n.c.) + S_K(axions) +S_K(anti-neutrinos) + etc.
\end{eqnarray}
In general, the variable $T$ may be dependent both on the detector-type and the type of its source, $T = T(S_i, j)$. Nevertheless, the assumption accepted here is the one that is already applied by Bahcall and Ulrich [54] and [6]. Bludman et al. [6] used $T_c$ as a phenomenological parameter representing different conceivable solar models in their equations for $S_{Cl}$, $S_{Kam}$ and $S_{Ga}$. This one-parameter allowance describes a quiet solar core with a temperature distribution similar to the SSM, therefore it leads to an SSM-like solution of the standard neutrino flux equations. Here and in the followings the temperatures are normalised to their SSM values. But let us calculate straight on this contribution from the standard neutrino flux equations, which describe the pp,CNO neutrino fluxes. 

One could think that there is no reason to introduce the temperature dependence to the neutrino flux equations since in that case we could loose the model independence. Nevertheless, the temperature dependency does not include solar model dependency as it involves only nuclear physics in the generalised approach presented here, because we did not apply the SSM luminosity constraint since it is based on an SSM dependent assumption which is counter-indicated. The model dependence of the previous calculations for the temperature dependent neutrino flux equations had arisen only through the SSM luminosity constraint that modified the original dependence of the neutrino fluxes [55]. Moreover, introducing the temperature dependence of the neutrino fluxes, it is possible the reach a physical insight which is able to reveal the basic causes of the neutrino problems. If we do not want to constrain ourselves to the SSM dependency, we have to abandon the SSM luminosity constraint, as it preserves the basic constraining assumptions about the types of nuclear reactions occurring in the solar core.  

An essential point in my calculations is that I have to use the temperature dependence proper in the case when the luminosity is not constrained by the SSM luminosity constraint, because another type of energy source is also present. The SSM luminosity constraint and the resulting composition and density readjustments, together with the radial extension of the different sources of neutrinos modify this temperature dependence. The largest effect arises in the temperature dependence of the $pp$ flux: $\Phi_1 \propto T^{-1/2} $ for the SSM luminosity constraint (see the results of the Monte-Carlo simulations of Bahcall, Ulrich [54]), but $\Phi_1 \propto T^4$ without the SSM luminosity constraint. Inserting the temperature-dependence of the individual neutrino fluxes for the case when the solar luminosity is not constrained by the usual assumption behind the SSM [55] into the chlorine-equation, we got the temperature dependent chlorine equation
\begin{eqnarray}
S_C(T) = a_{C1}T_{C,0}^4 \Phi_1(SSM) + a_{C7}T_{C,0}^{11.5} \Phi_7(SSM) + a_{C8}T_{C,0}^{24.5} \Phi_8(SSM)	
\end{eqnarray}
Similarly, the temperature-dependent gallium-equation will take the form:
\begin{eqnarray}
S_G(T) = a_{G1}T_{G,0}^4 \Phi_1(SSM)+ a_{G7}T_{G,0}^{11.5} \Phi_7(SSM) + a_{G8}T_{G,0}^{24.5} \Phi_8(SSM), 
\end{eqnarray}
and the SuperKamiokande equation shows that
\begin{eqnarray}
T_{K}=(\Phi_B(obs)/\Phi_B(SSM))^{1/24.5},
\end{eqnarray}
where $T_{i,0}$ refers to the neutrino-temperature of the solar core, seen by 
the $i$-type neutrino-detector, without including the hot bubbles (and their hot channels in which the bubbles move upwards). From these equations it is easy to derive the different neutrino-temperatures. With $\Phi_{pp}(SSM) = 5.94 \times 10^{10} cm^{-2}s^{-1}$, $\Phi_{Be}(SSM)=4.80 \times 10^9 cm^{-2}s^{-1}$, $\Phi_B(SSM)=5.15 \times 10^6 cm^{-2}s^{-1}$, $\Phi_B(SK,obs)=2.44 \times 10^6 cm^{-2}s^{-1}$, $S_C=2.56$, $S_G=72.2$ [14], $\Phi_1(SSM) = 5.95 \times 10^{10} cm^{-2}s^{-1}$, $\Phi_7(SSM) = 0.594 \times 10^{10} cm^{-2}s^{-1}$ and $\Phi_8(SSM) = 0.000515 \times 10^{10} cm^{-2}s^{-1}$. With these values, the chlorine neutrino-temperature $T_{Cl} \simeq 0.949 T(SSM)$, the gallium neutrino-temperature is $T_{Ga} \simeq 0.922 T(SSM)$ and the SuperKamiokande neutrino-temperature is $T_{SK} \simeq 0.970 T(SSM)$. The neutrino flux equations are highly sensitive to the value of the temperature. Assuming that the actual Sun follows a standard solar model but with a different central temperature, the above result shows that the different neutrino detectors see different temperatures. This result can be interpreted only with the conjecture that the different neutrino detectors show sensitivities different from the one expected from the standard solar model, i.e. some reactions produce neutrinos which is not taken into account into the standard solar model, and/or that they are sensitive to different types of non-pp,CNO reactions. Let us explore the consequences of this conjecture. 

%
%
\section{Discussion}
How to interpret these results? I suggest that the result $T_{Ga} < T_{Cl} < T_{SK}$ indicates that the chlorine and water detectors see some additional processes (or they have larger than expected sensitivities to some additional processes) to the ones included in the SSM. In this way, $T_{Ga} < T_{Cl}$ indicates that the chlorine detector has a larger sensitivity to a certain kind of non-pp,CNO neutrino in a certain energy interval. Accepting this conjecture, I am led to a result that intermediate or high energy neutrinos are produced by new type nuclear reactions in the Sun, and these contribute to the Homestake with a larger percentage than to the Gallex. 
To demonstrate the higher sensitivity of the Homestake in the intermediate ($\simeq 1 MeV$) and larger energies, it is enough to remember that in the Gallex flux the $pp$ neutrinos dominate, and the relative sensitivities of the $pp+pep, Be+CNO, B$ neutrinos in the Gallex to the Homestake show the ratios $\Phi_p(Ga:Cl) \simeq 333$, $\Phi_7(Ga:Cl) \simeq 32$, and $\Phi_8(Ga:Cl) \simeq 2.2$. 

Moreover, the result
$T_{Cl} < T_{SK}$ indicates a larger sensitivity of SuperKamiokande than Homestake to the non-pp,CNO reactions. While this may be right, since the SuperKamiokande is not sensitive to the neutrinos with energies smaller than 6,5 MeV, and so the relative weight of high-energy neutrinos is larger at it, another factor could work better to fulfil this task. This is that an additional neutrino flux is present at the SuperKamiokande, to which the chlorine detector is not sensitive at all. Plausibly enough, this can be produced by neutral currents, fluxes of muon (and perhaps tau) neutrinos. If we ignore at present the yet hypothetical MSW effect to be at work here, because it is not compatible with the other neutrino problems with three flavours, we would need a source that is able to generate muon (tau) neutrinos in the Sun. To generate muon neutrinos, it is necessary a high temperature (the estimated temperature is around $10^{11} K$), and high density for a significant amount of muon-neutrino flux. This temperature is just the one, which is calculated for the hot bubbles [13]. Moreover, the hot bubbles would be able to generate high-energy axions, and only the SK detector is sensitive to detect axions [56]. The contribution of the anti-neutrinos may be present also only in the SK data, but it seems not to be significant, as being less than $5.2 \%$ [57].

The result obtained by the above calculations gives expectation values for the future neutrino measurements different from the MSW calculations. For example, the dynamic solar model (DSM) suggest a $R_{Be} = \Phi_{Be}(observed)/\Phi_{Be}(SSM) \leq 39 \%$ depletion of the beryllium-neutrino flux, and $R_B = \Phi_B(observed)/\Phi_B(SSM) \leq 47 \%$ depletion for the boron-neutrino flux. These relative depletions refers to the "quiet solar core", to the part of the core without the runaway "hot bubbles" regions. Of course, the non-pp,CNO runaway (or "bubble") term produce an additional increase in the high energy neutrino spectrum. 
The differences between these depletions are not so significant as for the small-angle MSW solution, where $R_{Be} \simeq 0$ and $R_B \simeq 40 \%$ ([58], [59], [17], [22]). On the other hand, the large-angle MSW solution shows a nearly constant depletion above $1MeV$, $R(>1MeV) \simeq 0.2$ (see Fig. 16, in [1]). Therefore, the large angle solution of the MSW effect is not consistent with the preferential high-energy enhancement of the neutrino spectra ([60]-[62], [20]). If the mechanism suggested here - the functioning of a new nuclear reaction channel - works in the real Sun, then the Borexino has to measure a $ \Phi_b< R_{Be}(DSM) < 0.39 + \Phi_b$, for $0 < T < 0.922$. The derived results suggest values close to the upper limit. With T=0.922, the DSM prediction to Borexino is $\Phi_{Be}(DSM) = \Phi_{Be}(SSM) \times 0.39 + \Phi_b$ (here $\Phi_b$ is the bubble- and/or the non-pp,CNO neutrino flux), while the small-mixing-angle MSW solution would suggest a value close to zero. In this way, the derived predictions offer a possibility to distinguish between the case in which the MSW effect dominates $\Phi_{Be}(obs.) \simeq 0$, the other case in which a hybrid MSW+DSM mechanism works $0 < \Phi_{Be}(obs.) < 0.39$ and the third case in which the DSM mechanism works alone $\Phi_{Be}(obs.) \simeq 0.39$, or larger. Moreover, the spectrum above $5MeV$ should show a significant enhancement towards the larger energies. The dynamic solar model allows a larger place for $\Phi_{Be}$ because it works with a lower value for the standard neutrino fluxes at SuperKamiokande. Lowering the $pp$ neutrino flux, more place remains to the $Be$ neutrinos as well. Regarding the future measurements of SNO, the prediction of DSM is $\Phi_{\nu_e}(SNO, DSM) \simeq 0.16 \times \Phi_{\nu_e}(SNO,SSM) + \Phi_{\nu_e}(bubbles)$. Therefore, the charged current $CC(DSM) \simeq CC(SSM) \times 0.16 + CC(bubbles)$, and the neutral currents $NC(DSM) \simeq 1/3 \ CC(SSM) \times 0.16 + NC(bubbles)$.
  
If the Ga neutrino-temperature is $T_{Ga} \simeq 0.922 T(SSM)$, this means that the pp luminosity of the Sun is only $L_{pp} \simeq 72 \% L(SSM)$. The remaining part of the solar luminosity should be produced by the hot bubbles, 
$L_b \simeq 28 \% L(SSM)$. The new type nuclear reactions proceeding in the bubbles (and possibly in the microinstabilities) should also produce neutrinos, and this additional neutrino-production, $\Phi_b$ should generate the surplus terms in the chlorine and water Cherenkov detectors as well. At present, I was not able to determine which reactions would proceed in the bubbles, and so it is not possible to determine the accompanying neutrino production as well. Nevertheless, it is plausible that at that high temperature such nuclear reactions occur as at nova-explosions or other types of stellar explosions. Admittedly, these could be rapid hydrogen-burning reactions, explosive CNO cycle, and also nuclear reactions producing heat but not neutrinos, like e.g. the explosive triple-alpha reaction ([63]-[65]). At present, I remark that the calculated bubble luminosity ($ \simeq 28 \%$) may be easily consistent with the calculated non-pp,CNO neutrino fluxes $\Delta S_{Cl} = S_C(T_{Cl}=0.949) - S_C(T_{Ga}=0.922) \simeq 1.04$, $\Delta S_{Cl}/S_{Cl} \simeq 41 \%$, and $\Delta S_{SK} = S_{SK}(T_{SK}=0.970) - S_{SK}(T_{Ga}=0.922) \simeq 1.74$, $\Delta S_{SK}/S_{SK} \simeq 71 \%$.
 
The relation between the non-pp,CNO neutrino fluxes and the non-pp,CNO luminosity, together with the relation between the pp,CNO neutrino fluxes and the relevant pp,CNO luminosity, would stand to the place of the abandoned and overspecified solar luminosity constraint (this would be the generalised luminosity constraint). The above results are in complete agreement with the conclusion of Hata, Bludman and Langacker [17], namely: "We conclude that at least one of our original assumptions are wrong, either (1) Some mechanism other than the $pp$ and the $CNO$ chains generates the solar luminosity, or the Sun is not in quasi-static equilibrium, (2) The neutrino energy spectrum is distorted by some mechanism such as the MSW effect; (3) Either the Kamiokande or Homestake result is grossly wrong." These conclusions are concretised here to the following statements: (1) a non-pp,CNO energy source is present in the solar core, and the Sun is not in a thermodynamic equilibrium, (2) this non-pp,CNO source distorts the standard neutrino energy spectrum, and perhaps the MSW effect also contributes to the spectrum distortion (3) The Homestake, Gallex and SuperKamiokande results contains a term arising from the non-pp,CNO source,
which has the largest contribution to the SuperKamiokande, less to the Homestake, and the smallest to the Gallex. The results presented here suggest that the beryllium neutrino flux is lower than expected by the SSM because the neutrino temperatures (as measured by the different neutrino detectors) are lower than the expected SSM-value. The main reason is that because a non-pp,CNO energy source is also present in the solar core, the quiet SSM-like solar core has a lower neutrino temperature. Nevertheless, if the SSM electron neutrinos do take part in neutrino oscillation, the oscillation would lead to another factor which would depress the intermediate energy neutrinos, besides the apparent "cooling". 

One may think that the suggested mechanism could solve the solar neutrino problems, but new problems arose: the problem of the apparent "cooling" of the solar core, and the problem how the dynamic Sun is consistent with the helioseismic measurements. Although these questions would lead to another field, which does not necessarily belong to the present topic, let me outline some preliminary considerations. One thing is that the seismic temperatures and the neutrino temperatures do not necessarily wear the same values. The presence of an explosive energy source decouples the neutrino fluxes and temperatures from the seismic temperatures. Gavryusev [66] pointed out, that it is not possible to deduce directly the central temperature from the solar seismological data. Solar model calculations did show that the sound speed is an average property of the whole star and cannot be connected in any way to an "average temperature". The sound speed, as deduced from solar oscillations, is an "averaged sound speed" and it is a very stable value defined by global solar parameters (mass, radius, luminosity). Even significant changes in the inner solar model structure do not change it much. 

On the other hand, we can pay attention to the fact that the energy produced in the solar core do not necessarily pours into thermal energy, as other, non-thermal forms of energy may also be produced, like e.g. energy of magnetic fields. The production of magnetic fields can significantly compensate the change in the sound speed related to the lower temperature, as the presence of magnetic fields may accelerate the propagation of sound waves with the inclusion of magnetosonic and Alfven magnetohydrodynamical waves. 

The continuously present microinstabilities should produce a temperature distribution with a double character, as part of ions may posses higher energies. Their densities may be much lower than the respective ions closer to the standard thermodynamic equilibrium, and so they may affect and compensate the sound speed in a subtle way. Recent calculations of the non-maxwellian character of the energy distribution of particles in the solar core ([67], and more references therein) indicate that the non-maxwellian character leads to lowering the SSM neutrino fluxes and, at the same time, produces higher central temperatures. This effect may also compensate for the lowering of the sound speed by the lowering of central temperature. 

At the same time, an approach specially developed using helioseismic data input instead of the luminosity constraint, the seismic solar model indicates a most likely solar luminosity around $0.8 L_{Sun}$([33], Figs. 7-10), which leads to a seismological temperature lower than its SSM counterpart, $\Delta T \simeq 6 \%$. On the other hand, as Bludman et al. [6] pointed out, the production of high energy $^8B$ neutrinos and intermediate energy $^7Be$ neutrinos depends very sensitively on the solar temperature in the innermost $5 \%$ of the Sun's radius. In the region below $0.2$ solar radius the actual helioseismic datasets do not seem to offer reliable results ([69], see also [70], [50]).

%
%
\section{Conclusions}
A new, really model independent inequality is found between the neutrino detector rates. This inequality shows that neutrinos are detected in the neutrino detectors from a source besides the pp,CNO fluxes which are described by the standard neutrino flux equations. It is shown that first principle physics indicate thermonuclear runaways in stellar cores. These runaways may produce high-energy electron, muon and tau neutrinos and high-energy axions. The SuperKamiokande detects these particles. I derived how the generalised neutrino flux equations determine the contribution of the non-pp,CNO term in the neutrino detectors. The non-pp,CNO photon and neutrino luminosity seem to be consistent. It is argued that at present heioseismic data are contradicting below $0.2$ solar radius. Moreover, it seems that one need more careful analysis and more physical inputs to interpret properly the helioseismic data in the study of the innermost structure of the solar core in detail, since it is sensitive to the complex conditions present in the solar core. These include not only to the factors included in the standard solar model, but also to magnetic fields, nonequilibrium thermodynamics, and all the different manifestations of the energy production occurring in the solar core. And since together these sum up to the solar luminosity, the helioseismic data may allow cooler than standard Sun as well, because the thermal energy contains only a part of the total produced energy. Predictions of the dynamic solar core model are presented for the Borexino and SNO measurements that can distinguish between the cases when the MSW effect is dominant, the hybrid MSW+DSM mechanisms works, or the DSM mechanism dominates. 

The solution of the model independent neutrino flux equations strongly suggests that s new type of energy production mechanism is present in the solar core. 
The non-pp,CNO reactions are suggested to contribute to the production of intermediate, and, preferentially, of high energy neutrinos. Therefore, they are able to distort the solar neutrino spectrum in the way as it is indicated in [20], [62], without invoking new neutrino physics. The higher depletion of the intermediate energy neutrinos arises as a consequence of the lower than standard neutrino temperatures. The preferential enhancement in the high-energy region of the neutrino spectra is interpreted as enhanced by the contribution of thermonuclear runaways produced in micro- and macro-instabilities. 

The indicated presence of a non-pp,CNO energy source in the solar core - if it will be confirmed - will have a huge significance in our understanding of the Sun, the stars (and the neutrinos). This subtle and compact phenomena turns the Sun from a simple gaseous mass being in hydrostatic balance to a complex and dynamic system being far from the thermodynamic equilibrium. This complex, dynamic Sun ceases to be a closed system, because its energy production is partly regulated by tiny outer influences like planetary tides. This subtle dynamics is possibly related to stellar activity and variability. Modifying the participation of the MSW effect in the solar neutrino problem, the dynamic energy source has a role in the physics of neutrino mass and oscillation.
An achievement of the suggested dynamic solar model is that it may help to solve the physical and astrophysical neutrino problems without the introduction of sterile neutrinos, and, possibly, it may improve the bad fit of the MSW effect[39]. 
%
%
\section{Acknowledgements}
The work is supported by the Hungarian Scientific Research Foundation
OTKA under No. T 014224.

%
%
\section{Appendix-the thermal stability of stars. Excerpt from the book of Zeldovich, Blinnikov, Sakura}
The arguments for thermal instability of stellar cores of Zeldovich, Blinnikov and Sakura (1981) goes as following:
"Thermal equilibrium is determined by the equality of the speed of the energy liberation and the energy dissipation. The system is stable thermally, when to a small perturbation of the temperature these processes change in a way to decrease the initial perturbations. Let us assume, that in the centre hydrogen burns, and the liberated energy is dissipated through heat conduction. What happens for a small change of the temperature? From the diffusion equation of radiation one can get that the dissipation of heat depends on the physical parameters in the following way: $L_{ff}^- \sim RT^{7.5}/\rho^2$, when the 
Cramers opacity law is valid, and $L_c^- \sim RT^4/\rho$ in case of Compton scattering. The $R$ radius, $T$ averaged temperature and the $\rho$ density are independent from each other, as $M \propto R^3 \rho, T \propto GM/R$. Using these formulas, we got for the heat conduction the following formula
\begin{eqnarray}
L_{ff}^- \sim M^{a_1} \rho^{1/6} \sim M^{a_2} T^{1/2}
\end{eqnarray}
\begin{eqnarray}
L_c^- \sim M^{b_1} \rho^0 \sim M^{b_2}T^0
\end{eqnarray}
What is the situation if we introduce a heat surplus $L^+$? We can write that
\begin{eqnarray}
L_{pp}^+ = M \rho T^4 \simeq M^cT^7
\end{eqnarray}
for the pp cycle, while
\begin{eqnarray}
L_{CNO}^+ = M \rho T^{15} \sim M^d T^{18}
\end{eqnarray}
for the CNO cycle. In heat equilibrium, $L^+ = L^-$ at fixed temperature. The question is that this equilibrium is stable or not. At the first sight it seems that a tiny rise of the temperature leads to the increase of the nuclear energy liberation, and then this increasement rises the temperature further, so that the process leads to the explosion of the star...But the stars are stable anyhow. Where is the error in the above consideration? It is that we did not take into account the negative heat capacity of the star as a whole.  In the general case the change of temperature in the function of time is given by the equation
\begin{eqnarray}
cdT/dt = L^+ - L^-
\end{eqnarray}
where $c$ is the heat capacity of the system. Around the $T_0$ equilibrium point for small perturbations the $L^+$ and $L^-$ expressions may be given in terms of powers of $(T-T_0)$. Keeping only the first order quantities,
\begin{eqnarray}
cd(T-T_0)/dt = L^+(T_0) + dL^+/dT (T-T_0) - L^-(T_0) - dL^-/dT(T-T_0)
\end{eqnarray}
from which we can derive, using $L^+(T_0) = L^-(T_0)$, that
\begin{eqnarray}
cd/dt(T-T_0) = (dL^+/dT - dL^-/dT)(T-T_0)
\end{eqnarray}
From this it is clear that in case of $dL^+/dT > dL^-/dT$ and for $c > 0$ the small perturbations lead to instability:
\begin{eqnarray}
T-T_0 = const. \times exp[(dL^+/dT - dL^-/dT)/c]t.
\end{eqnarray}
The total energy $E$ of the star, i.e. the sum of the $U$ gravitational energy and the $W = C_vMT$ thermal energy will be negative. By the virial theorem
\begin{eqnarray}
U = -2W
\end{eqnarray}
\begin{eqnarray}
E = U + W = - W < 0.
\end{eqnarray}
Therefore the heat capacity of the star $c = dE/dT = -c_vM <0$. and so the small perturbations will decay and the star will be stable.
These (stabilisation) considerations are sound only 1.) when the star expands (contracts) as a whole during the increase of the small perturbations, and 2.) when the W thermal energy content is proportional to the temperature, i.e. when the stellar material is not degenerated. When these conditions are not present, instability may arise."
\eject

\end{document}